\documentstyle[12pt,floats,epsf,aps]{revtex}
 
\begin{document}

\newcommand{\beq}{\begin{equation}}
\newcommand{\eeq}{\end{equation}}
\title
{\bf On the origin of the radial flow in low energy heavy ion reactions } 

\author{\bf Regina Nebauer$^{*+}$  and J\"org Aichelin$^*$} 

\address{$^*$SUBATECH \\
Laboratoire de Physique Subatomique et des Technologies Associ\'ees \\
UMR Universit\`e de Nantes, IN2P3/CNRS, Ecole des Mines de Nantes\\
4, rue Alfred Kastler 
F-44070 Nantes Cedex 03, France.\\
$^+$ Universit\"at Rostock, Rostock, Germany}

\maketitle

\begin{abstract}
The average transverse energy  of nucleons and intermediate mass fragments 
observed in the heavy ion reaction Xe(50A~MeV)+Sn shows the same linear increase as a 
function of their mass as observed in heavy ion collisions up to the highest energies 
available today and fits well into the systematics. At higher energies this
observation has been interpreted as a sign of a strong radial flow in an 
otherwise thermalized system. Investigating the reaction with Quantum Molecular
dynamics simulations we find in between 50A MeV and 200A MeV a change in the
reaction mechanism. At 50A MeV the apparent radial flow is merely 
caused by an in-plane flow and Coulomb repulsion. The average transverse 
fragment energy does not 
change in the course of the reaction and is equal to the initial fragment energy
due to the Fermi motion.  
At 200A MeV, there are two kinds of fragments: those formed from spectator matter and those 
from the center of the reaction. There the transverse energy is caused by the pressure
from the compressed nuclear matter. In both cases we observe a binary event structure,
even in central collisions. This demonstrates as well the non thermal character 
of the reaction. The actual process which leads to multifragmentation is rather
complex and is discussed in detail.
\end{abstract}

\section{Introduction}

It is known since long that for almost all particles observed in heavy ion
reactions between $30A~MeV$ and $200~A.GeV$  
the transverse kinetic energy spectra have a
Maxwell-Boltzmann form, predicted for an emission from an equilibrated source.
However, the apparent temperature of the spectra and hence the average 
kinetic energy of the particles is quite different for different hadrons and
fragments and increases with increasing mass and increasing energy. This
observation seemed to exclude an identification of the apparent temperature 
with the real temperature of the system.

Recently it has been conjectured \cite{pbm,mar,reis} that at all energies 
between $50A~MeV$ and $200A~GeV$ the assumption of a strong radial flow can 
reconcile the mass dependence of the apparent temperature with 
thermodynamics.  At relativistic
and ultra-relativistic energies this has been inferred by comparing transverse
pion, kaon and proton spectra \cite{pbm}. At energies below
$500A~MeV$ the lever arm is still larger because one can include the 
intermediate
mass fragments (IMF's) of masses in between 2 and 10 \cite{mar,reis}, emitted
at mid rapidity, to separate radial flow and temperature. The deviations in forward 
and backward direction are usually interpreted as preequilibrium emission.

This observation has renewed the interest in the thermal analysis of heavy ion
reactions in the high as well as in the low energy heavy ion community
\cite{mar,her}. It revealed a very large value for
the radial flow velocity (up to 40\% of the speed of light) and a very peculiar 
beam energy dependence  
\cite{her} which 
has been discussed as a possible sign for a 
transition from a hadronic phase to a Quark Gluon Plasma phase.

Especially at low beam energies the observed 
value of the radial velocity in units of the speed of light of
$.05 \le \beta_r / c \le 0.2$ \cite{mar} (the exact value depends on the event
selection) is hard to understand in terms of physical processes because  
none of the other observables present evidence that the system is sufficiently
compressed for attributing the radial flow to an equation of state effect.
The very moderate in-plane flow is negative (as in deep inelastic collisions)
and points therefore to an other  
origin than compression and subsequent release of the compressional energy,
in contradistinction  to higher energies where the in-plane flow is positive.
Nevertheless, the low energy points follow smoothly the above mentioned 
systematics.

This radial flow at 50A~MeV has been reproduced\cite{neb}
by simulations using the Quantum Molecular Dynamics (QMD) \cite{aic}
approach. The same is true at higher energies 
where QMD simulations have been performed by the FOPI collaboration 
\cite{reis}.

It is the purpose of this article to take advantage of this agreement and to
study the origin of the large increase of the transverse energy with the
mass and hence the origin of the radial flow.  Here we concentrate 
ourselves on the reaction at 50A MeV and study in detail how the fragments
are produced. The reaction mechanism at this low energy is quite different than
that at higher energies. To demonstrate this we present as well the same study
for a beam energy of 200A~MeV.
A detailed study of the radial flow at higher energies, its equation of state
dependence and its physical origin will be the subject of a
forthcoming publication \cite{har}.

For our study at $(50A~MeV)$
we use simulations of the reaction  $Xe + Sn$
which has recently been measured by the INDRA collaboration. The INDRA detector
at GANIL has been constructed to study
multifragmentation and therefore the angular coverage and the 
energy thresholds
have been chosen to be better than that of any other $4\pi$ detector 
elsewhere. Hence the data taken with this detector are most suitable 
to confirm or disprove the theories embedded in the simulation programs. 
A detailed comparison of our results with the experimental data will
be published elsewhere \cite{neb}. Here we mention that not only the mass 
dependence of the average kinetic energy but also the kinetic energy spectra
themselves are in reasonable agreement with experiment.

For details about the QMD approach we refer to reference \cite{aic}. In
this program the nucleons are represented by Gaussian wave packets with a
constant width. The time evolution of the centers of these wave packets
is given by Euler Lagrange equations derived from the Lagrangian of the system.
The nucleons have an effective charge of Z/A. The fragment distribution becomes
stable after 240 fm/c \cite{pur}. From this point on we employ a Coulomb 
trajectory
program for the fragments and nucleons until the Coulomb energy is released.
This second step is necessary at this low energy because a large fraction of
the kinetic energy of the fragments is due to the Coulomb energy.   

\pagebreak


\section{The radial flow at beam energies of 50A MeV}

A first general idea of the time evolution of the collision can be obtained 
from the time evolution of the density of the system.
If the maximal density is reached the nuclei have their maximal overlap, after
 that the system
expands and the density decreases. This permits to find the time scale of 
the reaction. 

The total density is 
the sum over all nucleons which are described by Gaussians:
\begin{equation}
\label{densdef}
\rho(\vec{r},t)
\propto\sum_{i=1}^{A}
e^{-\frac{(\vec{r}-\vec{r_i}(t))^2}{2{L}}}
\label{denga}
\end{equation}
The width of the Gaussians is $4L=4.33~fm^2$ and A is the number of nucleons
present in the system.

\begin{figure}[h]
\vspace{-3.3cm}
\epsfxsize=17.cm
$$
\epsfbox{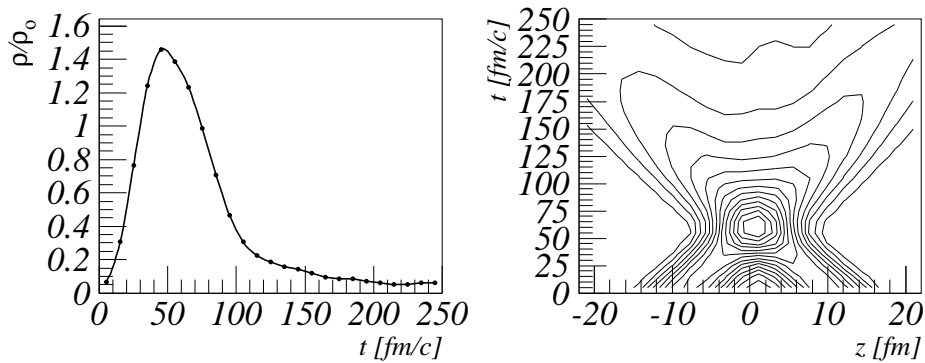}
$$
\vspace{-3cm}
\caption{\textit{Time evolution of the density (left) and of the density profile along
the beam (z) axis (right) for the system Xe(50A~MeV) + Sn ,b=3~fm.}}
\label{dic}
\end{figure}

In figure~\ref{dic}, left hand side, we plot the time evolution of the 
total density in the center of the reaction. 
The maximum density is obtained at $\approx 50~fm/c$, on 
the same time scale the
system expands and reaches at $120~fm/c$ a low density phase where the 
fragments do not interact anymore. 

On the right hand side of the same figure we display the density profile along the
beam (z) axis. Here we can follow the two nuclei. They occupy the same coordinate 
space at
$50..60~fm/c$. The system expands after $120~fm/c$. We find  
that this quasi-central ($b= 3~fm$) collision is semi transparent. Projectile
and target pass through each other without being seriously decelerated.
For $b=0~fm$ we get the same result.  
That binary character is confirmed by experiment:
In the center of mass system the experiment
shows 
a flat
angular distribution ($dN/dcos\theta_{cm}$) between
$60^\circ \leq \theta_{cm} \leq 120^\circ$ as well as 
a constant average kinetic 
energy  for fragments $Z\geq3$ even in central collisions. 
In forward and backward direction a strongly enhanced cross section 
is observed. The INDRA collaboration made use of this observation and
presented their data in two angular bins:
$60^o~\le~\theta_{CM}~\le~120^o$ (IMF's emitted in this angular range are called
mid-rapidity fragments (MRF's)) and $\theta_{CM}~<~60^o,\theta_{CM}~>~120^o$ called
projectile/target like fragments (PTF's) \cite{mar}.

How do the transverse fragment momenta reflect this passage through the other nucleus?
This is displayed in
figure 2 for a b = 3 fm reaction. Here we show the time evolution 
of the average transverse momentum of 
all fragments  with $ 5 \le A \le 10$ (left) and $ A > 10$ (right).  For the
dotted points the transverse momentum is represented with respect to the beam
direction, for the triangles with respect to the largest eigenvector of the
momentum tensor which is tilted by the flow angle $\theta_{flow}$.

\begin{figure}[h]
\vspace{-3.5cm}
\epsfxsize=16.cm
$$
\epsfbox{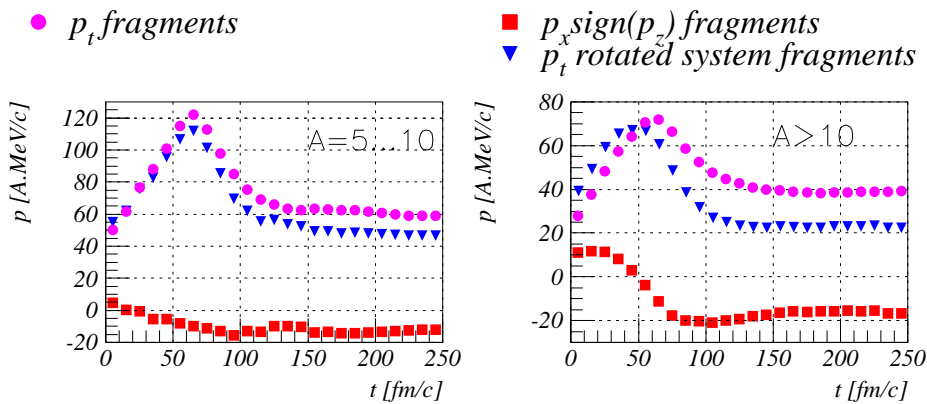}
$$
\vspace{-3cm}
\caption{\textit{Time evolution of the average transverse momentum
of the IMF's with respect to the beam direction (dots) and in the 
rotated system (by $\theta_{flow}$) (triangles) for 50A~MeV Xe + Sn, b = 3 fm.
The squares mark the time evolution of the in-plane flow.}} 
\label{ptrpzt}
\end{figure}

It is the first seminal result of this article that in the rotated system the
average transverse momentum is initially and finally the same. A very similar
result we obtain for b = 1 and b = 5 fm. During the reaction
the particles are accelerated in transverse direction but later they
feel a force into the opposite direction. The origin of this acceleration will
be discussed later. 
As squares we display the time evolution of the in-plane
flow ${1\over N_F}\sum_{N_F} sign(p^{cm}_z)p_x$. $N_F$ is the number of
fragments. We see that it remains moderate and is negative.

\begin{figure}[h]
\vspace{-2.5cm}
\epsfxsize=16.cm
$$
\epsfbox{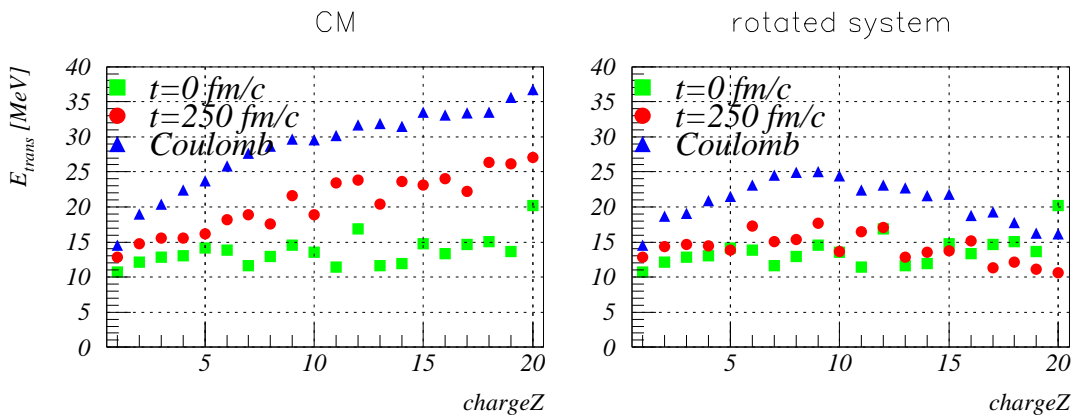}
$$
\vspace{-1.5cm}
\caption{\textit{Average transverse energy of the IMF's as a function of their
charge with respect to the beam direction (left) and in the 
rotated system (by $\theta_{flow}$) (right) for 50 A MeV Xe + Sn, b = 3 fm.}} 
\label{etx}
\end{figure}

The average  transverse energy of the fragments as a function of the fragment 
charge is 
displayed in fig. \ref{etx}. On the left hand side we display the transverse
energy with respect to the beam axis, on the right hand side with respect to
the rotated system. The mean value is presented after the initialization, after
250 fm/c, when the fragment distribution gets stable, and finally after the
mutual Coulomb repulsion has ceased. 

It is the second seminal result of this article that the average transverse 
fragment energy initially is about 14 MeV and independent of the fragment size. 
 As Goldhaber
\cite{gol} has pointed out many years ago this is expected if the
fragment formation is a fast process. In this case the momentum distribution of
the fragments is a convolution of the momentum  distribution of the
entrained nucleons and one expects for the
fragment momentum squared

$$P^2 =<(\sum_i^A p_i)^2> = {3k_{Fermi}^2 \over 5} A {N-A\over N -1}$$ 
where A is the number of  nucleons entrained in the fragment and N is the
number of nucleons of the disintegrating nucleus. Hence the fragment 
kinetic energy is  
$$ E_A = {3E_{Fermi} \over 5} {N-A \over N -1}$$
and  as well about 14 MeV in the QMD calculations. (Due to surface effects it
differs from the value of a Fermi gas in a square wall potential). 

How it can happen that the fragments pass the reaction zone without
being heated up and are finally observed at midrapidity or in PTF's  
we will investigate in the second part of this article.
The nucleons interact via the potential
\beq
V=-124\frac{\rho}{\rho_0} +70.5(\frac{\rho}{\rho_0})^2~[MeV]
\eeq
where the density is given by equation~\ref{denga} and $\rho_0$ is the normal
nuclear matter density.
In order to reveal the physics which drives the reaction we display the relative 
density of
those nucleons which are finally entrained in MRF's or PTF's as a function of time in
 the x-z plane
\beq
\rho^{\tiny MRF/PTF}_{rel}(x,z,t)={\rho_{\tiny MRF/PTF}(x,z,t) \over \rho_{total}(x,z,t)}
\eeq 
and superimpose the gradient of the potential in the x-z plane as arrows where
x is the direction of the impact parameter. 
For the sake of a clearer display we plot nucleons coming
from the projectile only. We start out with a discussion of the reaction 
at zero impact parameter then we continue with $b= 3 fm$ reactions.

\begin{figure}[h]
$$
\epsfxsize=16.cm
\epsfbox{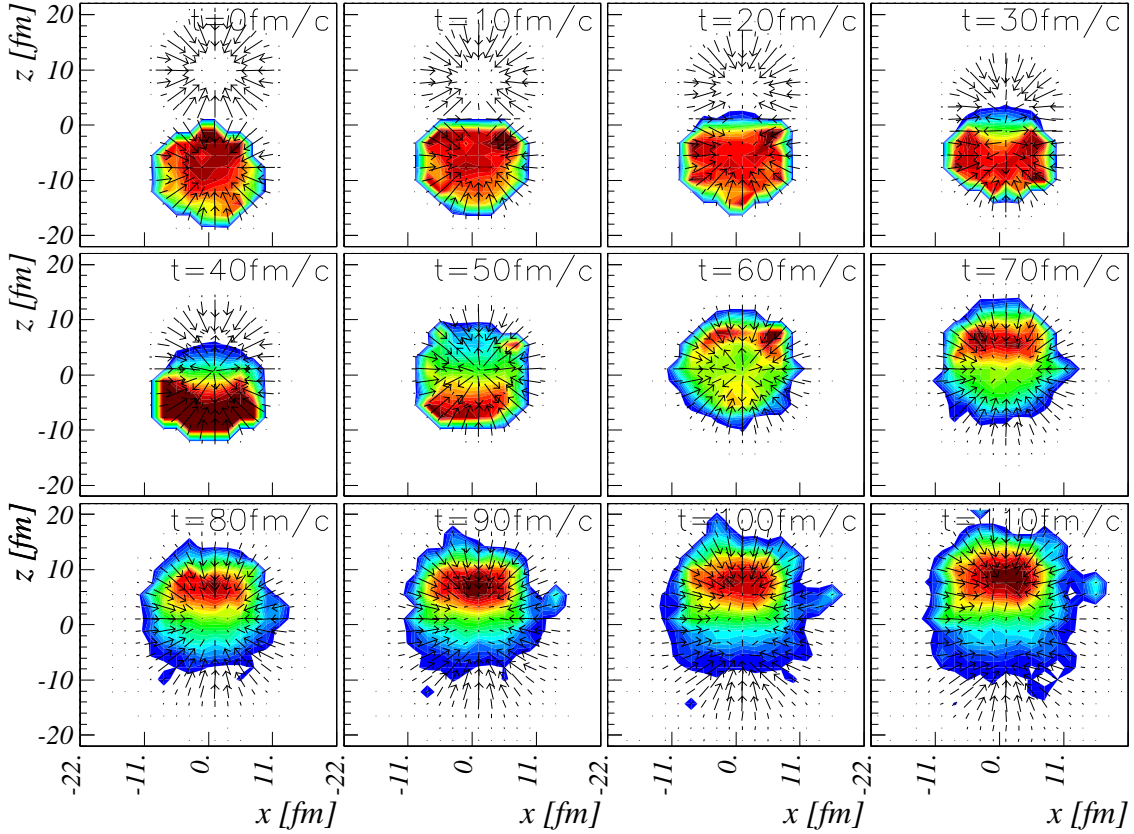}
$$
\vspace{-1.cm}
\caption{\textit{Reaction Xe(50A~MeV)+Sn. Movement of the nucleons finally emitted as IMF's in forward/backward
direction in the mean field potential for collisions at impact parameter $b=0~fm$. We 
display the fraction of these nucleons on the
total density (shadow) and the gradient of the potential (arrows) projected on the x-z
plane}}
\label{potgfg0}
\end{figure}

The motion of the nucleons in the potential of a nucleus is a sequence of acceleration
and deceleration. Nucleons on the surface are almost at rest, due to the density (and
thus the potential) gradient they become accelerated towards the center of the nucleus. 
They
reach their maximal momentum when they pass
the center of the nucleus, climb up the potential on the 
other side and are finally
at rest again when arriving at the surface. 
When a heavy ion collision occurs, the position of
the nucleons in the projectile or target determines whether they "feel"
the heavy ion collision right from the beginning or only when the
high density phase has already passed. We will show that the initial position 
of the entrained nucleons decides 
as
well whether the fragment is finally observed at mid rapidity or in forward/
backward direction.

In figure~\ref{potgfg0} we display the motion of the nucleons finally entrained in
PTF's for a reaction at  $b=0~fm$. The spatial distribution of those nucleons 
is almost identical with
that of all nucleons present in the projectile.
In the first step of the collision the nucleons move away from the target
into the yet unperturbed part of the projectile. When they arrive
at the back end of the projectile they invert the direction of their momenta. 
They are then accelerated in longitudinal direction towards the 
center of the reaction. When they arrive finally there the high density
zone has disappeared already.
Hence the nucleons pass the center without a larger change of their initial 
momentum. The initial correlations \cite{goss} among the nucleons which
finally form a fragment survive the reaction because
all potential gradients are small.

\begin{figure}[h]
$$
\epsfxsize=16.cm
\epsfbox{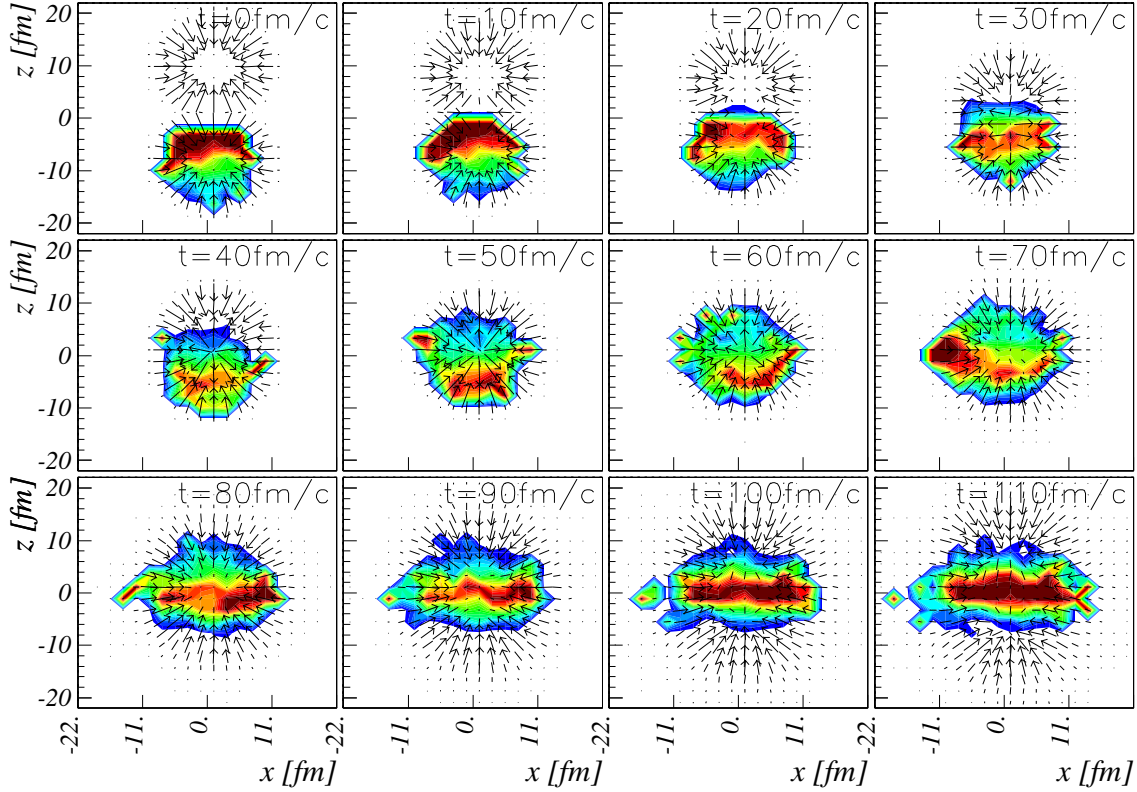}
$$
\vspace{-1.cm}
\caption{\textit{Reaction Xe(50A~MeV)+Sn. Movement of the nucleons finally emitted as 
IMF's in mid-rapidity in 
the
mean field potential for collisions at impact parameter $b=0~fm$. We display the
 fraction 
of these nucleons on the
total density (shadow) and the gradient of the potential (arrows) projected on the x-z
plane}}
\label{potgm0}
\end{figure}

Nucleons finally emitted as MRF's (fig. \ref{potgm0}) are strongly
located at the front end of the nuclei. These are the nucleons which 
climb the 
nuclear potential created by the higher density in the reaction zone and which are 
at rest before they are on the top of the potential wall. 
Due to their position they are involved in 
the collisions between projectile and target nucleons right from the beginning. 
Collisions support the
deceleration. (Later collisions are to a large extend Pauli suppressed.) 
They escape the barrier 
in transverse direction. 
As their (longitudinal as well as 
transverse)momentum
is quite small, the nucleons stay longer in the
center of the reaction what favours the mixing of projectile and target 
nucleons. 
When leaving the reaction zone the fragments become decelerated due to the
potential interaction with the rest of the system. This 
deceleration balances
the gain in energy due to the prior acceleration in transverse direction, 
although the
physics of both processes is rather independent.

\begin{figure}[h]
$$
\epsfxsize=16.cm
\epsfbox{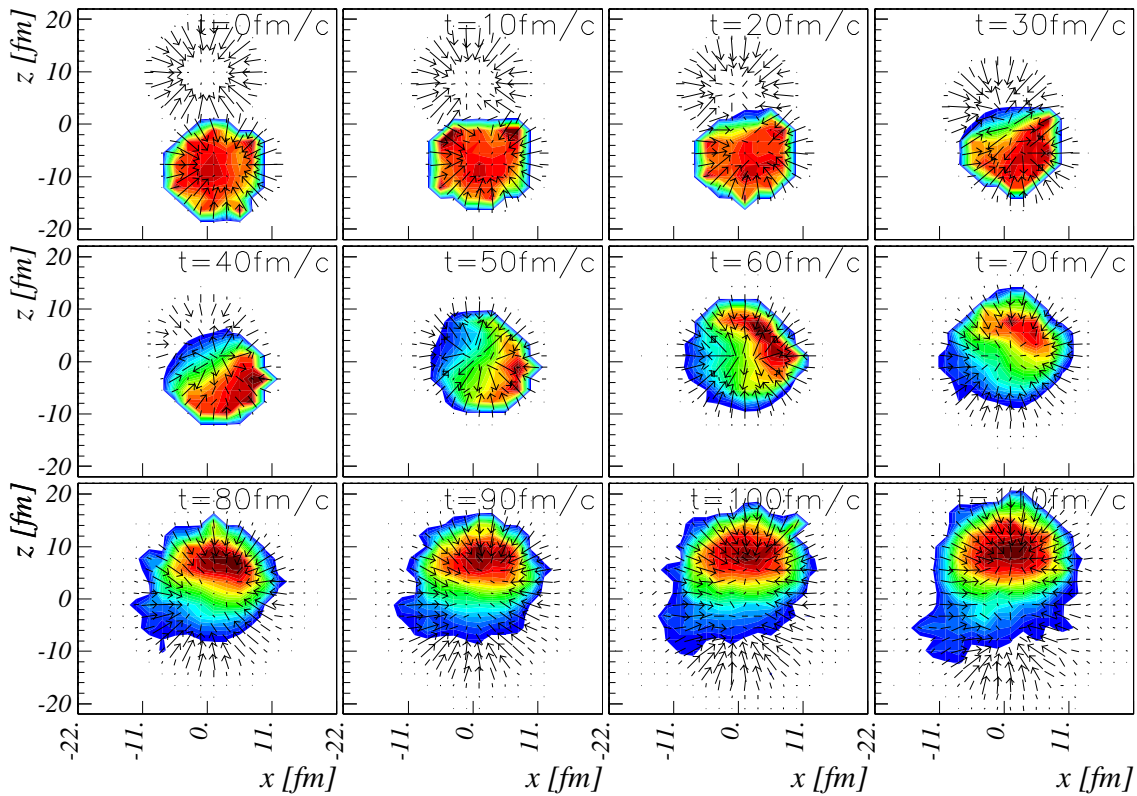}
$$
\vspace{-1.cm}
\caption{\textit{Reaction Xe(50A~MeV)+Sn. Movement of the nucleons finally emitted as IMF's in forward/backward
direction in the mean field potential for collisions at impact parameter $b=3~fm$. We 
display the fraction of these nucleons on the
total density (shadow) and the gradient of the potential (arrows) projected on the x-z
plane}}
\label{potgfg3}
\end{figure}

The same analysis for PTF's for $b=3~fm$ is displayed in figure \ref{potgfg3}. 
The general reaction mechanisms, even the time scales,
are the same as discussed before. In addition to the scenario at $b=0~fm$ 
asymmetry 
effects occur here. Up to $40~fm/c$ the scenario is the same as at
zero impact parameter. When the high density phase occurs, the nucleons take the line of
least resistance, i.e. they follow the minimum of the potential on the right hand
side (for the projectile nucleons, the target nucleons take the inverse direction on the
other side). The larger part of the nucleons pass the reaction center when the
potential barrier has disappeared. As already discussed for zero impact 
parameter case, the
nucleons emitted as fragments in forward/backward direction pass through the 
reaction zone without an important change of their initial momenta.

For the reaction Xe(50A~MeV)+Sn we could find two modes for the emission of fragments:

\begin{enumerate}
\item The Fragments are already "preformed" in the nuclei. The initial correlations
between the nucleons survive as they avoid the high density zone. For a finite impact
parameter a strong in-plane flow is observed. The fragments are quite big and emitted in
forward/backward direction.
\item Fragments emitted in midrapidity are smaller. They are formed by mixing projectile
and target nucleons. Due to their initially small longitudinal momentum they stay in the
high density zone and emitted in transverse direction. These fragments are rather small as
they are formed via collisions.
\end{enumerate}


\section{The radial flow at beam energies of 200A MeV}

In the reaction at 200 A~MeV the velocity of the nucleons due to 
internal motion and the velocity of the two colliding nuclei in the center of mass frame
are now significantly different. As a consequence we expect a change of the reaction
mechanism.

\begin{figure}[h]
\vspace{-3.5cm}
\epsfxsize=16.cm
$$
\epsfbox{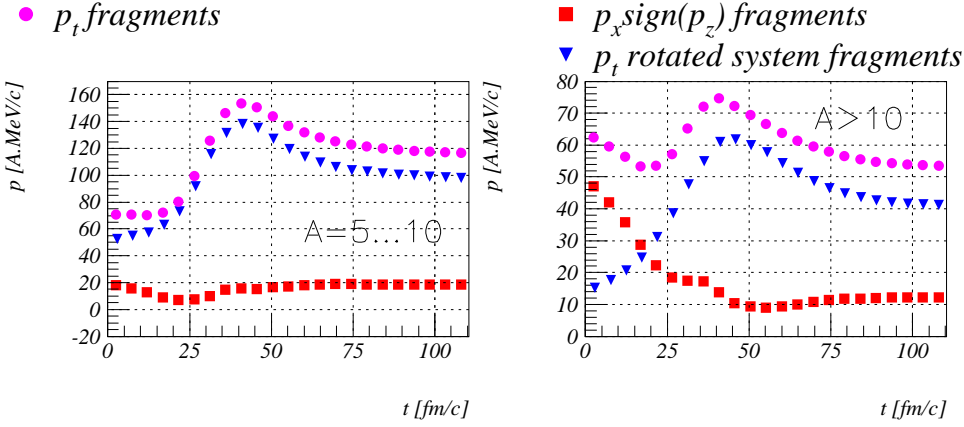}
$$
\vspace{-2cm}
\caption{\textit{Time evolution of the average transverse momentum
of the IMF's with respect to the beam direction (dots) and in the 
rotated system (by $\theta_{flow}$) (triangles) for 200A~MeV Xe + Sn, b = 3 fm.
The squares mark the time evolution of the in-plane flow.}} 
\label{ptt200}
\end{figure}

That this is indeed the case can already be seen from the time evolution of the 
transverse momenta of the fragments, figure~\ref{ptt200}. The collective directed in-plane
flow has the opposite sign as compared to 50A MeV. This is as well a sign for the 
change of the reaction mechanism which develops between 50A MeV and 200A MeV from a 
deep inelastic scenario 
with a negative flow angle to a bounce off type reaction with a positive flow angle.  

Despite of the higher beam 
energy the transverse acceleration of the light
fragments is delayed as compared to the reaction at 50A MeV and sets on only
at the maximum overlap of the two nuclei. Then they are
accelerated and remain with a high transverse momentum after the expansion.

The large fragments have a completely different time evolution as compared to
50A MeV.  First we see a deceleration in transverse direction of the
fragments. Then, when the highest density is reached this turns into an 
acceleration. Finally, as at lower energies, the 
fragments are decelerated again during the expansion of the system. Finally they have
a quite low transverse momentum.

\begin{figure}[h]
\vspace{-1cm}
\epsfxsize=16.cm
$$
\epsfbox{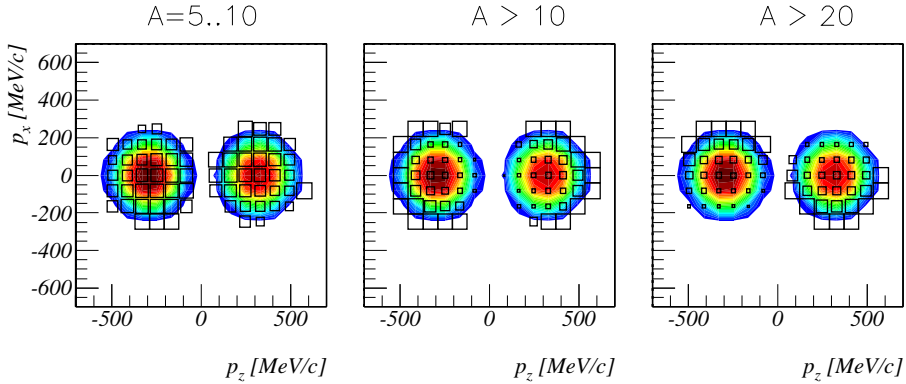}
$$
\vspace{-2cm}
\caption{\textit{Initial - final - state correlations in momentum space. We represent the 
positions of the nucleons at t=0 (color level) and superimpose as boxes the partition of
nucleons finally emitted as small (lhs), big (middle) or very big (rhs) fragments.}} 
\label{ifscg}
\end{figure}

To understand this behaviour it is useful to study the initial-final state correlations.
Particularly in momentum space we observe such strong correlations. In figure~\ref{ifscg} we
display the initial distribution of the nucleons at t=0 (color level) and superimpose
the relative fraction  of nucleons finally emitted in small (lhs) or big (middle, rhs) fragments.
Here we can clearly see that the nucleons finally emitted in heavy fragments have 
an initial momentum which points away from the reaction partner. This initial - final -
state correlation, already
allusively present at 50A MeV increases with the mass of the fragment and the beam energy
\cite{goss}. 

For nucleons finally
emitted in small fragments the correlations are less pronounced but with preference of momenta
which point into the reaction zone what favours the mixing of projectile and
target nucleons in the fragments. 
Nucleons emitted finally in small fragments are located in the center of the collision
partners and become only accelerated when a higher density is built up in the center of the
reaction. Following the density gradient they develop a radial flow. Finally they
disentangle form the system. This costs energy and therefore the momentum decreases.
We expect a constant acceleration of all fragment nucleons and therefore 
a final transverse energy which increases linearly with the fragment mass. 
The light fragments show only a small in-plane-flow which is positive in contrast to the
reaction at the lower energy.  In contradistinction to 50A MeV we observe
for the light fragments a genuine acceleration in transverse direction independent of the
system in which it is measured. At 200A MeV the density built up in the center of the reaction
is sufficiently high for a noticeable acceleration of the fragments.
We see that especially for the heavy fragments that the initial transverse momentum in 
the rotated system is quite different as compared to the beam system due to the initial
correlations in momentum space, discussed above. 

\begin{figure}[h]
\vspace{-1cm}
$$
\epsfxsize=16.cm
\epsfbox{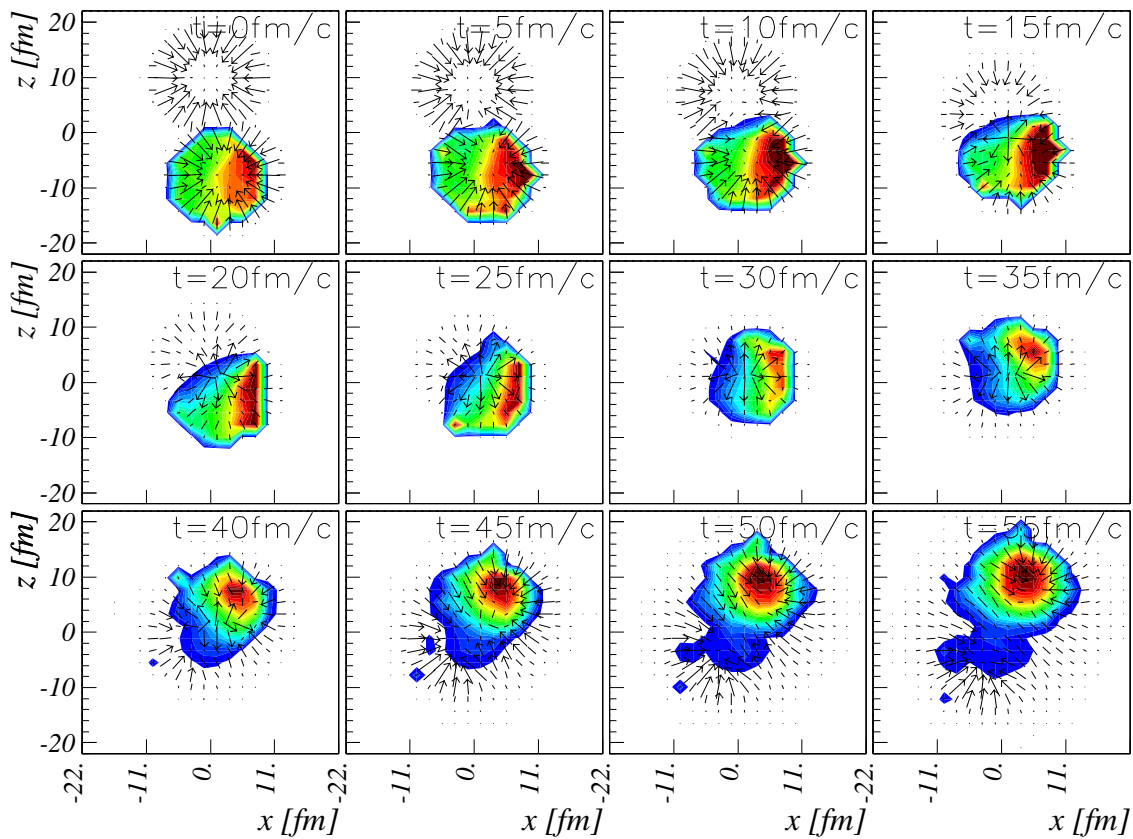}
$$
\vspace{-.5cm}
\caption{\textit{Reaction Xe(200A~MeV)+Sn. Movement of the nucleons finally emitted in
big fragments ($A>10$). We display the fraction of these nucleons on the
total density (shadow) and the gradient of the potential (arrows) projected on the x-z
plane}}
\label{potb200}
\end{figure}

For a better understanding of the time evolution of the momenta of the big fragments we 
study as at 50A~MeV the movement of their constituent nucleons in the nuclear potential, 
figure~\ref{potb200}. Here we display the gradient of the nuclear potential and the 
relative fraction of nucleons emitted in fragments $A>10$.  
 
In addition to their strong initial - final - state correlations in momentum space
the fragments have as well strong correlations in coordinate space. The big fragments 
are at t=0 located at the surface of the collisions partners and have
therefore an average density well below the normal nuclear matter density. At the beginning
of the reaction
they are accelerated towards the center of the reaction because the density is 
higher there.
This lowers the transverse momentum. When the high density at the center of the reaction is
built up, the acceleration changes sign. Thus we can already at 200A MeV and 3~fm 
impact parameter distinguish between participants and spectators.

Now we can draw up two competing processes of fragment formation at 200A~MeV: 

\begin{enumerate}
\item  Large fragments are formed from spectator matter. The entrained nucleons do not pass
the region of high density and therefore the acceleration in transverse direction is moderate.
For the entrained nucleons we observe as well strong initial - final - state correlations 
in momentum space. In other words, the more the nucleons avoid the high density zone the larger
is the change that they find themselves finally in big fragments. At 200A MeV the
probability that nucleons passes the high density zone without having a collision is well
reduced with respect to 50A MeV. Therefore any larger correlated part of nuclear matter which
passes the high density becomes destroyed by collisions and cannot form finally a fragment
anymore.

\item Small correlated parts of nuclear matter can still survive the passage through the high
density zone. That will be different at still higher energies. These
nucleons form small fragments mixing projectile and target
via collisions. The nucleons finally emitted in this fragments are close to 
the high density zone and accelerated due to the internal pressure. Their 
transverse energy is high and expected to be proportional to their mass (radial flow).
\end{enumerate}

\begin{figure}[h]
\vspace{-2.cm}
\epsfxsize=16.cm
$$
\epsfbox{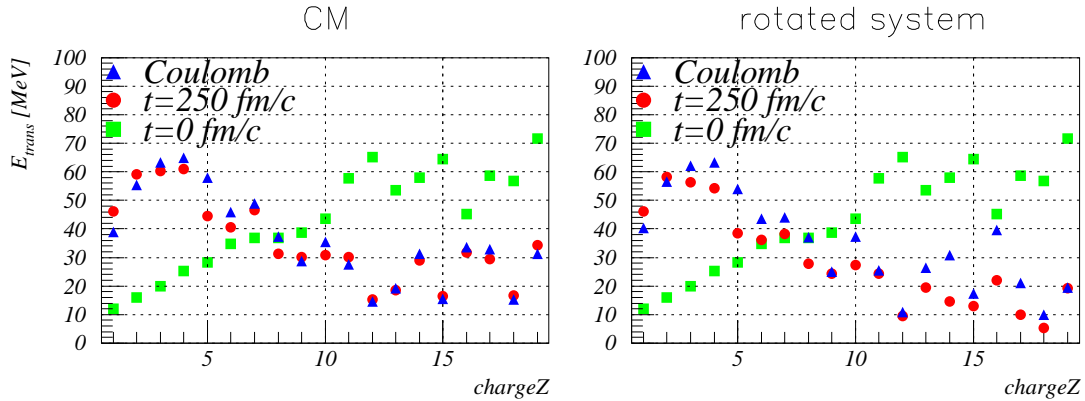}
$$
\vspace{-1.5cm}
\caption{\textit{Average transverse energy of fragments as a function of their
charge with respect to the beam direction (left) and in the 
rotated system (by $\theta_{flow}$) (right) for 200A~MeV Xe + Sn, b = 3 fm.}} 
\label{et200}
\end{figure}

These two competing processes allow  to understand 
the behaviour of the transverse energy as a function of the 
charge (figure~\ref{et200},lhs). At t=0, the beginning of the reaction, the strong 
initial correlations in
momentum space (high transverse momentum for big fragments)
cause a linear increase of the transverse energy with the charge of the
fragment. For small fragments the process 2 dominates, finally the
transverse energy of fragments is large and increases with increasing charge (mass) of the
fragment.
For big fragments the process 1 is dominant, the average transverse
energy is small. In between we observe a transition phase.
On the rhs of the same figure we display as in figure~\ref{etx} the transverse
energy of the fragments in the rotated systems which is almost identical with the left hand
side. Hence the radial energies are genuine and not artifacts due to a strong in-plane flow as
at 50A MeV. 


\section{Conclusion}

In conclusion we have found that the kinetic energy of the fragments observed in
the heavy ion reaction $Xe + Sn$ at $50A~MeV$ reflects the initial Fermi energy of 
the entrained nucleons. There is no hint that the system becomes equilibrated. Rather, in 
agreement with experiment,
even in central collisions the reaction is semi transparent. We find that the final
longitudinal momentum of the fragments depends strongly on the time point when the
entrained nucleons pass the center of the reaction.
Nucleons which are passing  early in the reaction are decelerated
due to collisions with the reaction partner and due to the strong potential gradient.
They are deviated into the transverse direction as their longitudinal momentum is not
 high enough to
overcome the potential barrier formed in the reaction center in the phase of
highest density.
 Nucleons which arrive later 
do not encounter a strong potential gradient anymore and pass the reaction center
freely. Thus, they keep almost their initial momentum. We find that
compressional effects have little influence on the final momentum of the 
fragments at this energy and the observed apparent radial flow is 
not real. The linear increase of the fragment 
kinetic energy with the mass for small fragments
finds its natural explication in terms of the initial Fermi motion,
the Coulomb barrier and a small in-plane flow. 

At 200A~MeV we found that the reaction mechanism has changed and approaches the
participant-spectator model as far as heavy fragments are concerned. Due to the smaller mean
free path heavy fragments cannot come anymore form the center of the reaction. Their
transverse energy remains therefore moderate and the in-plane flow weak.
Small fragments show a strong radial flow and have passed the center of the reaction 
where the forces are much stronger than at 50A MeV.

At higher energies fragments formed from of participants become more and more rare due to
the shorter mean free path and due to stronger force gradients. Hence fragment formation
of spectator matter will dominate. Initial - final - state correlations in momentum space become
stronger and collisions start to contribute in addition to the radial
flow. How these different process influence the final radial flow and its dependence on the
nuclear equation of state is presently under investigation \cite{har}.


\begin{thebibliography}{99}
\bibitem{pbm} P. Braun-Munzinger et al, Phys. Lett. {\bf B344} (1995) 43 and
              Phys. Lett. {\bf 365} (1996) 1
\bibitem{mar} N. Marie, PhD Thesis , Universit\'e de Caen, France
\bibitem{reis}W. Reisdorf, Nucl. Phys. {\bf A612}  (1997) 493, 
              G. Poggi et al., Nucl Phys {\bf A586}
             (1995) 755, B. Hong et al. nucl-ex/9707001
\bibitem{her} N. Herrmann, Nucl. Phys. {\bf A610} (1996) 49c, 
\bibitem{neb} R. Nebauer at al., nucl-th/9810008
\bibitem{aic} J. Aichelin, Phys. Rep. {\bf 202} (1991) 233.
\bibitem{har} C. Hartnack et al., to be published  
\bibitem{pur} R. Puri et al., Phys. Rev. {\bf C54} (1996) R28
\bibitem{gol} A.S. Goldhaber Phys. Lett. {\bf B53} (1974) 306
\bibitem{goss} P.B. Gossiaux and J. Aichelin, Phys. Rev. {\bf C56} (1997) 2109 
\end{thebibliography}
\end{document}